# An agent-based negotiation model and its implementation in Repast

S. Bai[1]

**Abstract**

We propose an agent-based model, MNegoti, for simulating multilateral negotiation process, which can be naturally employed in group decision support system. This model can also be applied to any use case in which negotiation is involved, in order to simulate the negotiation process. In this report, we discuss the implementation of the MNegoti model on the basis of the agent-based simulation platform, Repast Simphony. It is worth pointing out that this model can be used to create a java module for any use of agent-based negotiation simulation.

## 1. Introduction

Negotiation has been an important field of study within organizational behavior and management science. It can be defined as a discussion between two or more parties with conflicting interests aiming to reach an agreement [1]. In recent decade, automated negotiation has stimulated growing interests of researchers and many works have been done, e.g. [2].

Agent-based simulation has been a species of computation, growing up alongside the maturation of computer technology. Agent-based model (ABM) consists of a collection of autonomous and heterogeneous agents interacting with other agents and the environment by exchanging their attributes/preferences or states, and the agents have access to past and current values of their own state variables. Agents make decisions using both prescribed rules and analytical functions [3] and [4]. ABM becomes a major bottom-up tool that has been employed into many research fields. The Repast Suite is a family of advanced, free, and open source agent-based modeling and simulation platforms that have been under continuous development for over 15 years. Repast Simphony is a richly interactive and easy-to-learn Java-based modeling system that is designed for use on workstations and small computing clusters. Repast has found many interesting applications in diverse fields and most of the corresponding research works can be found in the website [5].

In this report, we propose an agent-based negotiation model, MNegoti, for describing multilateral negotiation process. This report is organized as follows. Essential features of the MNegoti model are listed in Section 2, and specific elements for implementing the MNegoti model in Repast are briefly discussed in Section 3. Finally, summary of the MNegoti model and possible future directions are given in Section 4.

## 2. MNegoti model

MNegoti model is an agent-based negotiation model which describes multilateral negotiation process and a concrete implementation of the MNegoti model was realized in one intelligent strategy evaluation

---

[1] Karlsruhe Institute of Technology, Institute for Thermal Energy Technology and Safety

E-Mail: shan.bai@kit.edu

system [6]. Two basic entities of this model are agent and meeting room. The conceptualizations of the model, e.g. time-dependent state variables and actions of these entities, are summarised in the following.

- ❖ Package: Topics
  - ➢ Class: Criteria for agent groups
    Criteria are introduced and each agent group can assign to its preference in a certain range, the preference obeying some distribution when considering one special criterion.
  - ➢ Class: Negotiation issues
    In [7], the negotiation issues are protection strategies in nuclear and radiological emergency management and preparedness.
  - ➢ Class: Negotiation strategies
    Negotiation strategies can be either dynamic or static. A good performance can be achieved by utilizing Strategy pattern in behavior software design pattern. Static strategies can also be implemented by retrieving existing strategies from data storages (e.g. warehouse).
- ❖ Package: Agents and environments
  - ➢ Class: Agent group
    - Identification, for example, the type, the name, or others which can identify the group uniquely.
    - Boundaries and distributions of preferences related to criteria. Boundaries can be realized in the data structure, e.g. matrix, which each row contains the lower and upper boundaries of the preference w.r.t. one criterion.
    - The number of the members.
  - ➢ Class: Agent
    - Group Identification: Usually, one agent belongs to one agent group.
    - Preferences whose values obey the probability distributions and fall within the range determined by the boundaries of preferences of the agent group to which the agent concerned belongs.
    - evaluate(): Evaluating negotiation issues by agent's preferences and other information, e.g. a kind of weight. This assessment can be implemented in diverse embodiments and used to define the utility of the agent.
    - @Watch (e.g. the other agents, the open state of meeting room) in order to schedule some method.
  - ➢ Class: Meeting room
    - open(): The meeting room is available under a certain condition for some agent to exchange information and negotiate. There are two embodiments for admitting the agents to the meeting room:
      o Set the conditions/agenda for the negotiation. Agents who watch the meeting room can check the conditions/agenda and those who satisfy these conditions or are interested in the agenda can enter into the meeting rooms.
      o Send directly the invitation to special agents.
    - negotiate(): Agents in the meeting room negotiate together, that is, a multilateral negotiation process.
      o Choose suitable negotiation protocol, for example, discussion in groups, among groups, etc.
      o Choose suitable negotiation strategies.
        ▪ Cooperative: trade-off, etc.
        ▪ Non-Cooperative: bid, etc.

Meeting room can be open many times because of the requirements from agents to negotiate about different agenda and so on. Moreover, one negotiation process can last one or more time steps ([7]) in meeting room. It can also be the situation that many meeting rooms are open and the agents attend negotiations with different agenda in those meeting rooms simultaneously.

## 3. MNegoti model in Repast

### 3.1 Context

Repast introduces a data structure, Context, which is a simple container based on set semantics. Any type of object can be put into a Context with the simple caveat that only one instance of any given object can be incorporated into the Context. From a modeling perspective, the Context represents an abstract population. The objects, referred to as proto-agents, in a Context are the population of a model. The Context is actually more of a proto-space. The Context provides the basic infrastructure to define a population and the interactions of that population without actually providing the implementations. As a proto-space, the Context holds proto-agents that have idealized behaviors, but the behaviors themselves cannot actually be realized until a structure is imposed on them. [8]

The entity, meeting room, in our work can be one generalisation of the Context because there are particular properties that a meeting room can be open and closed at each run time, the agents observe its state, and when it is closed, the results from the negotiation process are stored in a certain way in the meeting room which are useful for its next open.

### 3.2 Network projection

Repast introduces projections that specify the relationship between the agents in a given Context. Here we focus on network projections where the relationships (e.g. social, spatial) among the agents can be defined. A projection is attached to a particular Context and can be applicable to all of the agents in that Context. [8]

The entity, agent group, is regarded as beyond the network projection because it contains more information, for example, in the MNegoti model, the boundaries of preferences. One agent can get its own preference values from the preferences in its group by obeying a certain rule. Agent group can be implemented by Factory method pattern in software so as to create new agents.

### 3.3 Scheduling using watchers

The current version of Repast Simphony introduces one new scheduling approach, scheduling using watchers. Watcher queries are Boolean expressions, which can evaluate the watcher and the watchee with respect to each other and some projection or Context. Repast schedule can be stopped or paused by model code, either scheduled at predetermined time, or immediately via some event. Schedulers are implemented as secret event clock manager, where the clock, measured in "ticks", is incremented at the completion of the execution of all the actions at that clock tick. In sum, a watcher allows an agent to be notified of a state change in another agent and schedule an event to occur as a result. The watcher is set up using an annotation, but instead of using static times for the schedule parameters, the user specifies a query defining whom to watch and a query defining a trigger condition that must be met to execute the action. [8]

In the MNegoti model, the agents make use of this watcher mechanism to check the open conditions of meeting rooms on each tick and decide whether to enter this or another meeting room. Each meeting room will send invitations to specified agents if necessary in each tick. The negotiations can be processed when negotiators are ready. These actions from the agents and the meeting room shall be executed sequentially which can be performed in Repast, e.g. by setting priorities in @ScheduleMethod. Negotiation processes can last in one tick or many ticks depending on the negotiation protocols and strategies.

## 4. Discussion

This report introduces the architecture of the agent-based model for multilateral negotiations, MNegoti. More details including how to take preference values for agents, negotiation agenda, protocols, strategies and so on, can be found in [7]. There are still open questions about how to define independent criteria, extend agent and meeting room e.g. by adding proper actions and novel state variables, and so on. Furthermore, the MNegoti model can be implemented without using any feature of Repast.

MNegoti model has a wide range of prospective applications. It can be used naturally for group decision making as a computational implementation. Moreover, this model can be introduced into any use case in which negotiation is involved, in order to simulate the negotiation process. For example, for supply chain, negotiation issues may be quantities of products, the capabilities of transport, etc. On the other hand, it is worthwhile to consider further developments of the MNegoti model in view of different application scenarios, for example, to introduce various evaluation methods, specific negotiation agenda, protocols and strategies to be adopted on a case-by-case basis.